\documentclass{article}

\usepackage[english]{babel}

\usepackage[letterpaper,top=2cm,bottom=2cm,left=3cm,right=3cm,marginparwidth=1.75cm]{geometry}

\usepackage{amsfonts}
\usepackage{amssymb}
\usepackage{amsbsy}
\usepackage{amsthm}
\usepackage{titling}
\usepackage{tikz}
\usepackage{float}
\restylefloat{table}
\usetikzlibrary{shapes,arrows}
\usepackage{mathtools}
\usepackage{amsmath}
\usepackage{graphicx}
\usepackage[colorlinks=true, allcolors=blue]{hyperref}

\newtheorem{theorem}{Theorem}

\title{USING IMPORTANCE SAMPLING IN ESTIMATING WEAK DERIVATIVE}
\author{Cheng Jie\\ [12pt]
Department of Mathematics \\ 
University of Maryland \\
Glenn Martin institute of Technology\\
College Park,MD 20740, USA\\
\and
Micheal C.Fu\\[12pt]
Division of Operations Research \\
University of Maryland\\
Van Munching Hall\\
College Park, MD 20740, USA \\
}

\begin{document}
\maketitle

\begin{abstract}
In this paper we study simulation-based methods for estimating gradients in stochastic networks.
We derive a new method of calculating weak derivative estimator using importance sampling transform, and our method has less computational cost than the classical method. In the context of M/M/1 queueing network and stochastic activity network, we analytically show that our new method won't result in a great increase of sample variance of the estimators. 
Our numerical experiments show that under same simulation time, the new method can yield a narrower confidence interval of the true gradient than the classical one, suggesting that the new method is more competitive.
\end{abstract}

\section{Introduction}

Stochastic gradient estimation plays an important role in many fields, such as simulation optimization\cite{Hastings1970MonteCS} and the study of option price sensitivities (\cite{Fu2002OptimizationFS}). 
Suppose we are interested in 
\begin{equation}
\frac{d J(\theta)}{d \theta},
\end{equation}
where $J(\theta)$ is not directly available but instead the simulation model returns a randomized output $Y(X,\theta)$ such that
$$
J(\theta)=
E(Y(X,\theta)),
$$
where, $Y(X,\theta)$ can represent, for instance, the total cost time of stochastic activity network or the average waiting time in G/G/1 queue. Estimating the gradient in (1) can itself be a challenging task, which constitutes the subject of this paper.

A remarkable amount of stochastic gradient estimation methods have been developed in past few decades,and they can be essentially  classified into two main categories--- indirect or direct estimations.The most popular indirect gradient estimators method include finite difference and simultaneous perturbation. Direct estimators include score function(SF) method, infinite small perturbation analysis(IPA) and measure valued differentiation(weak derivative). We will write them respectively as SF, IPA and WD for the sake of simplicity in the rest of paper. 

Unlike SF and IFA, which are single-run estimators, weak derivatives often requires more computational time to simulate random variable $X^+$ and $X^-$ and calculate the difference of the values $Y(X^+)$ and $Y(X^-)$ [see section 2]. Such a disadvantage make WD estimator less widely used though it usually has less variance than single-run estimators.

The importance sampling method was originally developed for the purpose of variance reduction. However, it can been used in many cases other than variance reduction. For example, one can use it to study one distribution while simulating another(\cite{Owen2019ImportanceST, zhang2016construction, Regueiro2014a, zhang2015},). Inspired by the idea of importance sampling, we develop a new way to estimate weak derivative,called importance sampling weak derivative(ISWD) estimator, which can be get in a single-run.

However, since the importance sampling transform may yield a estimator with variance dramatically larger than the original one, 
one has to be very cautious of using that\cite{10.3389/fdata.2022.966982, jiecheng-thesis}. The remainder of this paper is devoted on discussing different cases where our new method is applicable and if so, whether it can outperform the classical way to estimate Weak Derivative.

\emph{Preview}.This paper has the following structure. Section 2 introduces the direct gradient estimation problem and presents weak derivative estimators. Section 3 introduces importance sampling and present our new way of calculating the WD estimator. 

\section{WEAK DERIVATIVE ESTIMATOR}

\subsection{Introduction}

\hspace{13pt} We denote our objective function to be $J(\Theta)$, the randomized output value to be $Y(\textbf{X})$, and the following equality holds: 
$$E\{Y(\textbf{X})\} =J(\Theta)$$

where $\textbf{X}$ denote the vector of input random variables of the stochastic network. 
$
\textbf{X}=
\{ 
X_i...X_N
\}
$, 
and 
$\Theta=\{\theta_1,...\theta_n\}$ 
denote the parameter associated with 
$\{X_1.....X_N\}$.

We are interested in the sensitivity of $\theta_i$ in our objective function \cite{LIN20181, Jie2021BiddingVC}, which can be expressed as

 \begin{equation}
 \frac{d E(Y(\textbf{X}))}{d \theta_i} =\frac{ d J(\Theta)   }  {  d \theta_i  }.
 \end{equation}
 
 Writing equation (2) in the integration form, we have

\begin{equation}
\frac{d E(Y(\textbf{X}))} {d \theta_i} 
= \int_{
\mathbb{R}
} 
Y(x_1,...x_N) 
f(x_1,...x_N;\theta_i) 
dx_1...dx_N
\end{equation}

 Suppose the r.v.s 
$ \{ 
X_i...X_N
\}$
are independent to each other, i.e

$$
f(x_1,...x_N,\Theta) 
=
\prod
_1^N
f(x_i;\theta_i)
$$
Where $f(x_1,...x_N;\Theta)$ represents the joint density function of 
$
\{ 
X_i...X_N
\}
$

The derivative can then be written as 
\begin{equation}
\frac{d E(Y(\textbf{X}))} {d \theta_i} 
=
 \int_{
\mathbb{R}
} 
Y(x_1,...x_N)  
\frac
{
\partial f
_i
(x_i;\theta_i)
}
{
\partial \theta_i
}
f_{-1}(x_1....x_N)
dx_1...dx_N
\end{equation}
Where 
$
f_{-1}(x_1....x_N)
$
denote the joint density function of all the r.v.s except 
$X_i$.
\\

Observe the fact that when the derivative 
$\partial f
i
(x_i;\theta_i)
/\partial \theta_i
$
can be decomposed weakly to the subtraction of two nonnegative functions, that is
\begin{equation}
\frac
{
\partial f
_i
(x_i;\theta_i)
}
{
\partial \theta_i
}
=
c_i(\theta_i)
(
f_i^+(x_i;\theta_i)
-
f_i^-(x_i;\theta_i)
),
\end{equation}
Such weak decomposition of derivative can be seen by the property that 

$$
\int_{-\infty}^x 
\frac
{
\partial f_i(x_i; \theta_i)
}
{
\partial \theta_i
}
=
\frac{\partial F_i(x_i; \theta_i)}{\partial \theta_i}
=c_i(\theta_i)\int_{-\infty}^{x_i}
f^+(t;\theta_i)-f^-(t;\theta_i)  dt  
$$
which is justified by the absolute continuity property of the derivative $\partial F(x_i;\theta)/ \partial \theta$.
\\

According to The weak decomposition of derivative\cite{Jie2018StochasticOI} of density function leads to the following relationship: 

\begin{equation}
\frac{d E(Y(\textbf{X}))} {d \theta_i} 
=
 \int_{
\mathbb{R}
} 
Y(x_1,...x_N)  
c_i(\theta_i)
(
f_i^+
(x_i;\theta_i)
-
f_i^-
(x_i;\theta_i)
)
f_{-1}(x_1....x_N)
dx_1...dx_N
\end{equation}
For simplicity, we denote (5) as
\begin{equation}
\frac{d E(Y(\textbf{X}))} {d \theta_i} = 
\sum_{i=1}^n c_i(\theta_i) 
E
(
Y(X_1,...X_i^+,...X_n)- 
Y(X_1,...X_i^-,...X_n)
)
\end{equation}
with 
$X_i^+\sim f_i^+(x_i;\theta_i)$, 
$X_i^-\sim f_i^-(x_i;\theta_i)$
and $X_i^+$, $X_i^-$ are all independent of 
r.vs $(X_1,...X_{i-1},X_{i+1},..X_n)$.And a typical weak derivative estimator for the gradient can be written as
\begin{equation}
\{
\sum
_1^
{N}
c_i(\theta_i)
(Y(X_1,.X_i^+..X_
{N}
)
- 
Y(X_1,.X_i^-..X_
{N}
)
)
\}
\end{equation}

In the special case where $X_is$ are i.i.d, $c_i(\theta_i)s$ and $\theta_i s$ are identical, denoted by$\theta$ and $c(\theta)$, and 
we let $J(\theta)$ represents our objective function,and we are interested in estimating

 \begin{equation}
 \frac{d E(Y(\textbf{X}))}{d \theta} =\frac{ d J(\theta)   }  {  d \theta  }.
 \end{equation}

The WD estimator becomes
\begin{equation}
c(\theta)
\{
\sum
_1^
{N}
(Y(X_1,.X_i^+..X_
{N}
)
- 
Y(X_1,.X_i^-..X_
{N}
)
)
\}
\end{equation}

\subsection{Motivational Examples}
\begin{itemize}

\item \textbf{Gaussian distribution \cite{Winkelbauer2012MomentsAA}}

Suppose $X\sim \mathcal{N}(\theta,\sigma^2)$, 
with density function 

$$
\phi_{\theta,\sigma}(x)
=
\frac{1}
{
\sqrt{2\pi} \sigma^2} e^{-(x-\theta)^2/2\sigma^2
}
$$
According to Fu(06), we can decompose the derivative weakly to the following:  

$X^{+} \sim \theta+Wei(2,\frac{1}{2\sigma^2})$, and $X^{-} \sim \theta-Wei(2,\frac{1}{2\sigma^2}) $. 
\\
\item \textbf{Gamma Distribution \cite{la2016cumulative}}

Denote 
$X\sim  Gamma  (\alpha,\theta)$  
with density equal to 
$$
f_{\theta,\alpha}(x)=
\frac
{
\theta
^{-\alpha} 
x^
{ \alpha-1 }
e^{-x/\theta}
}
{
\Gamma
(
\alpha
)
}
$$
one choice of weak derivative decomposition can be:

$X^+\sim gamma(\alpha+1,\theta)$, 
$X^- \sim gamma(\alpha, \theta)$. 
\\
\item \textbf{Exponential Distribution}

Consider $X\sim exp(\theta)$ , a special case of Gamma distribution with density: 
$$
f_{\theta}(x)=
\frac
{1}
{\theta}
e^
{
-
\frac
{1}
{\theta}
x
}
$$
it's weak derivative decomposition can be: 
$X^+\sim erlang(2,\theta)$
,
$X^- \sim exp(\theta)$

\end{itemize}

\section{IMPORTANCE SAMPLING WEAK DERIVATIVE ESTIMATOR}

\subsection{Review of importance sampling}
\hspace{0.15in} 
 
 Suppose that our problem here is to find 
 $$\mu_p=E_p(Y(X))=\int_{\mathcal{D}}Y(x)p(x)dx$$ 
where $p$ 
is a probability density function of X with support in 
$\mathcal{D}$. 
If q is a positive probability density function with support larger than
$\mathcal{D}$, then

\begin{equation}
\mu_p=\int_{\mathcal{D}}Y(x)p(x)dx
=\int_{\mathcal{D}}\frac{Y(x)p(x)}{q(x)} q(x) dx
=\mathbb{E}_q(\frac{Y(X)p(X)}{q(X)})
\end{equation}

\noindent We denote $\mu_q$ for the value $E_q(Y(X)p(X)/q(X))$ in the rest of this section.

Given that we have n samples with density $q$, denoted by 
$\{X_1...X_n\}$,
the \textbf{importance sampling estimator} of 
$\mu_p=\mathbb{E}_p(Y(X))$ 
is 
\begin{equation}
 \hat{\mu}_p=\frac{1}{n}\sum_{i=1}^n\frac{Y(X_i)p(X_i)}{q(X_i)}, X_i \sim q.
\end{equation}

The variance of importance sampling transform is given in the theorem below without proof: 

\begin{theorem} 
 Let $\hat{\mu}_p$ be given by (2) where $\mu=\int_{\mathcal{D}} Y(x)p(x) dx$ and $q(x)>0$ whenever $Y(x)p(x)\neq 0$. Then 
$\mathbb{E}_q(\hat{\mu_q})=\mu$, and $Var_q(\hat{\mu}_q)=\frac{\sigma^2_p}{n}$ where 
\begin{equation}
 \begin{split}
\sigma^2_q &=\int_{\mathcal{D}} \frac{(Y(x)p(x))^2}{q(x)} dx-\mu^2 \\
           &=\int_{\mathcal{D}}\frac{(Y(x)p(x)-\mu q(x))^2}{q(x)} dx
\end{split}
\end{equation}

\end{theorem}

According to the theorem above, one can naturally find the estimator for the variance $\sigma_q^2$ 
$$
 \hat{\sigma^2_q}=\frac{1}{n}\sum_{i=1}^n(\frac{Y(X_i)p(X_i)}{q(X_i)}-\hat{\mu}_q)^2
$$

Therefore an approximate $95\% $ CI for $\mu$ is $\hat{\mu_q} \pm 1.96\hat{\sigma^2_q}/\sqrt{n}$
\\

$\bullet$ \textbf{Remark:} At this time the adjustment factor $p(x)/q(x)$ is called the \textbf{likelihood ratio}, The distribution q is the 
\textbf{importance distribution} and p is the \textbf{nominal distribution}.
In the above case the importance distribution $q(x)$ doesn't have to be positive everywhere.
It's sufficient that we have $q(x)>0$ whenever $Y(x)p(x)\neq 0$.

\subsection{Application in estimating Weak Derivative}
Consider estimating weak derivative described in section 2, under the assumption all the input r.v.s $\{X_1,...X_n\}$ are independent to each other, with associated $\{\theta_1....\theta_n\}$.

Observe that 
$$
\frac
{
dE(Y(X_1,.....X_n))
}
{
d\theta_i
}
=
c_i(\theta_i)
(E
(
Y(X_1..X_i^+..X_n)-Y( X_1..X_i^-..X_n)
)
)
$$
Therefore, based on the equation above, we can develop a modified weak derivative estimator(will call it ISWD in the following paper) of 
$
\frac
{
dE(Y(\theta_i))
}
{
d\theta_i
}
$
to be
$$
c_i(\theta_i)
Y(X_1...X_i...X_n)
(
\frac
{
f^+_i(X_i)
}
{
f_i(X_i)
}
-
\frac
{
f^-_i(X_i)
}
{
f_i(X_i)
}
) 
$$

Observe that if we assume all $X_is$ are i.i.d, the expression in (7) can also be written as 

\begin{equation}
\sum_{i=1}^n c_i(\theta_i) 
E
(
Y(X_1,...X_i^+,...X_n)- 
Y(X_1,...X_i^-,...X_n)
)
=
\sum_{i=1}^n 
c_i(\theta)
E
(
Y(X_1,...X_i,...X_n)
(
\frac{f_i^+(X_i)}{f_i(X_i)}
-
\frac{f_i^-(X_i)}{f_i(X_i)}
)
)
\end{equation}
And based on (11), a sample of Modified weak derivative estimator for 
$\frac{dE(Y(\textbf{X}))}{d \theta}$ shall be

\begin{equation}
\sum_{i=1}^n
c(\theta)
(
Y(X_1,...X_i,...X_n)
(
\frac
{
f_i^+(X_i)
}
{
f_i(X_i)
}
-
\frac
{
f_i^-(X_i)
}
{
f_i(X_i)
}
)
)
\end{equation}

\subsection {ISWD estimator in M/M/1 FCFS queue}
\hspace{0.15in}

\noindent $\bullet$ \textbf{Notations:}

Let 
$
T_N
(
X_1,..X_N,A_1,...A_N
)
$ denote the system time of the $N_{th}$ customer, with 
$X_1,..X_N$
represent the r.v.s of the service time\cite{zhu2020high} and 
$A_1,...A_N$
represent the r.v.s of the inter arrival time. Denote $\theta$ the mean of service time, and we assume all 
$X_1,..X_N$ are i.i.d exponentially distributed. Denote $f$ to be the density of $X_i$, and $f^+$, $f^-$ been respectively, the density from weak decomposition of $\partial f(\theta)/\partial \theta$.

We are interested in estimating the sensitivity\cite{zhu2022clustering, ZHAO2018619, Senseney2017, Zhang2015fracture} of expectation of Nth customer's system time w.r.t the mean service time, which is 
$$
\frac{d E(T_N(X_1,..A_N))}{d\theta}
$$

\noindent $\bullet$ \textbf{Typical weak derivative estimator\cite{Flynn2019ASP}:}

Notice that since all $X_i s$ are i.i.d, $c_i(\theta)$ are identical to each other, we denote them as $c(\theta)$.
The sensitivity of system time w.r.t the mean service time can be written as
\begin{equation}
\frac{d E(T_N(X_1,..A_N))}{d\theta}\\
=c(\theta) \sum_{i=1}^N E(T_N(X_1,..X_i^+..X_N,A_1,...A_N)-T_N(X_1,..X_i^-...X_N,A_1,..A_N)).
\end{equation}

Where 
$X_i^+$ 
and 
$X_i^-$ are random variables whose densities come from weak derivative decomposition of density of $\partial f/\partial \theta$. 
\\

The sample estimator for the sensitivity becomes 
$$
c(\theta)
\sum_{i=1}^N
(
T_N(X_1,..X_i^+..X_N,A_1,...A_N)
-
T_N(X_1,..X_i^-...X_N,A_1,..A_N)
)
$$

\noindent $\bullet$ \textbf{ISWD estimator:}

Inspired by the framework of importance sampling, one can instead estimate the expectations 
through ratio of the density function 
$f^+/f$ and 
$f^-/f $, i.e,

\begin{equation}
\frac{d E(T_N(X_1,..A_N))}{d\theta}
=
\sum_{i=1}^N
c(\theta)
(
E(T_N(X_1,...A_N)
\frac{f^+
(X_i)}
{f
(X_i)})-
E(T_N(X_1,...A_N)
\frac
{f^-
(X_i)}
{f(X_i)})
)
\end{equation}

This time, the \textbf{estimator} is been reduced to 

$$c(\theta) \sum_{i=1}^N T(A_1,...X_N)\frac{f^+(X_i)}{f(X_i)}- T(A_1,...X_N)\frac{f^-(X_i)}{f(X_i)}$$

\noindent $\bullet$ \textbf{Computation time comparison:}

The computational time for the typical estimator include the following three parts: 
\begin{enumerate}
\item simulate $X_i^{+}$ and $X_i^{-}$, 
\item calculate
$T_N(X_1,...X_i^+,..X_N,A_1,..A_N)$
 and 
$T_N(X_1,..,X^-..X_N,A_1,...A_N)$ 
accordingly 
\item subtracting the above two values. 
\end{enumerate}
Among the three parts, calculating $T_N$ consumes the most and especially when N is large, it often take a long time. 

\textbf{However}, the modified estimator will only consume as much as the 3rd part of typical estimator.

\bibliographystyle{unsrt}
\bibliography{sample}

\end{document}